\documentclass{svproc}
\usepackage{url}

\usepackage{graphicx}
\usepackage{color}
\usepackage{amsmath}
\usepackage{amssymb}
\usepackage{nicefrac}
\usepackage{tabularx}
\usepackage{booktabs} 
\usepackage{colortbl}
\usepackage{verbatim} 
\usepackage{enumitem}  
\usepackage{xspace} 
\usepackage{tikz}
\usepackage{nicefrac}
\usepackage{relsize}
\usepackage{soul}
\usepackage[backend=bibtex,style=trad-plain]{biblatex}
\usepackage{wrapfig}

\newcommand{\pT}{$p_{\mbox{\tiny T}}$\xspace}
\newcommand{\sNN}{$\sqrt{s_{\mbox{\tiny NN}}}$\xspace}
\newcommand{\sNNF}{\sNN~=~2.76~TeV\xspace}
\newcommand{\sNNMa}{\sNN~=~5.02~TeV\xspace}

\newcommand{\GeVc}{GeV/$c$\xspace}
\newcommand{\vtwo}{$v_2$\xspace}

\newcommand{\piz}{$\pi^{0}$\,}

\newcommand{\Rg}{$R_{\gamma}$\xspace}

\begin{document}
\mainmatter              
\title{Direct photon and light neutral meson production in the era of precision physics \\ at the LHC}
\titlerunning{Direct photons and neutral mesons measured with ALICE}  
%
\author{Meike Charlotte Danisch on behalf of the ALICE Collaboration}
\authorrunning{Meike Charlotte Danisch on behalf of the ALICE Collaboration} 
%
%
\institute{Physikalisches Institut, Universit\"at Heidelberg, Heidelberg, Germany\\
\email{meike.charlotte.danisch@cern.ch}}

\maketitle              

\begin{abstract}

In these proceedings we present the latest results from ALICE on direct photon and light neutral meson production in p--Pb and Pb--Pb collisions.
The direct photon excess ratio \Rg in different charged particle multiplicity classes of p--Pb collisions at \sNNMa is shown.
In addition, we present the direct photon elliptic flow coefficient \vtwo in central and semicentral events of Pb--Pb collisions at \sNNF.
An outlook on ongoing and future measurements is given.

\keywords{direct photons, electromagnetic probes, heavy-ion collisions, quark-gluon plasma}  
\end{abstract}

\section{Introduction}

Direct photons, being defined as photons not originating from hadron decays, are a valuable tool to investigate the space-time evolution of the medium created in heavy-ion collisions.
The measured elliptic flow coefficient of photons reflects the momentum anisotropy of the source, convoluted with the photon emission rate, integrated over time.
Direct photons can also contribute to the effort of testing bulk effects in high multiplicity collisions of small systems.
 
There are different sources of photons in heavy-ion collisions. 
Prompt photons are created in initial hard scatterings. 
In addition, we expect thermal photons from the QGP and the hadron gas phases, which are sensitive to the medium temperature \cite{1}.
There will also be photons from hadron decays, which are the vast majority of all photons. The latter originate mostly from decays of the neutral mesons \piz and $\eta$ into two photons.
Therefore we need to measure their spectra precisely in order to be able to obtain the excess of direct photons over this decay photon backgound. 
Due to the different shapes of transverse momentum (\pT) distributions of prompt and thermal photons (power-law and close-to-exponential respectively) they can be distinguisted on a statistical basis.
Prompt photons dominate direct photons at high \pT$\gtrsim4\,$\GeVc and thermal photons at low  \pT$\lesssim3\,$\GeVc.  
Therefore, a direct photon excess at low \pT can be interpreted as thermal photon signal.
So far it was assumed that thermal photons are relevant only in AA collisions but after collective effects have been observed also in high-multiplicity pp and p--A collisions (see for example \cite{2}) we can question this assumption.


\section{Direct photon \Rg in p--Pb collisions at \sNNMa}

Observation of a direct photon signal implies that the ratio of inclusive ($\gamma_{\text{inc}}$) over decay photons ($\gamma_{\text{dec}}$) is larger than one. In order to eliminate parts of the uncertainty, we define the double ratio as 
$R_{\gamma} = \frac{\gamma_{\text{inc}}}{\gamma_{\text{dec}}} \equiv \frac{\gamma_{\text{inc}}}{\pi^{0,\text{p}}}$/ $\frac{\gamma_{\text{dec}}}{\pi^{0,\text{p}}}$, where $\pi^{0,\text{p}}$ stands for the parametrisation of the measured \piz spectrum.
In the analysis presented here, inclusive photons are measured with three different techniques: PCM, EMCal and PHOS \cite{3}. 
The EMCal is a sampling calorimeter composed of alternating layers of lead and plastic scintillators.
It is placed at a radius of $R=4.3$\,m from the beam and covers a range of $\Delta \varphi = 100^{\circ}$ in azimuthal angle and $|\eta| < 0.7$  in pseudorapidity. The cell size is about $6 \times 6\,\text{cm}^2$.
\begin{wrapfigure}[25]{l}{0.48\textwidth}   
  \begin{center} 
  \includegraphics[width=0.48\textwidth,trim=0 30 0 80,keepaspectratio=true]{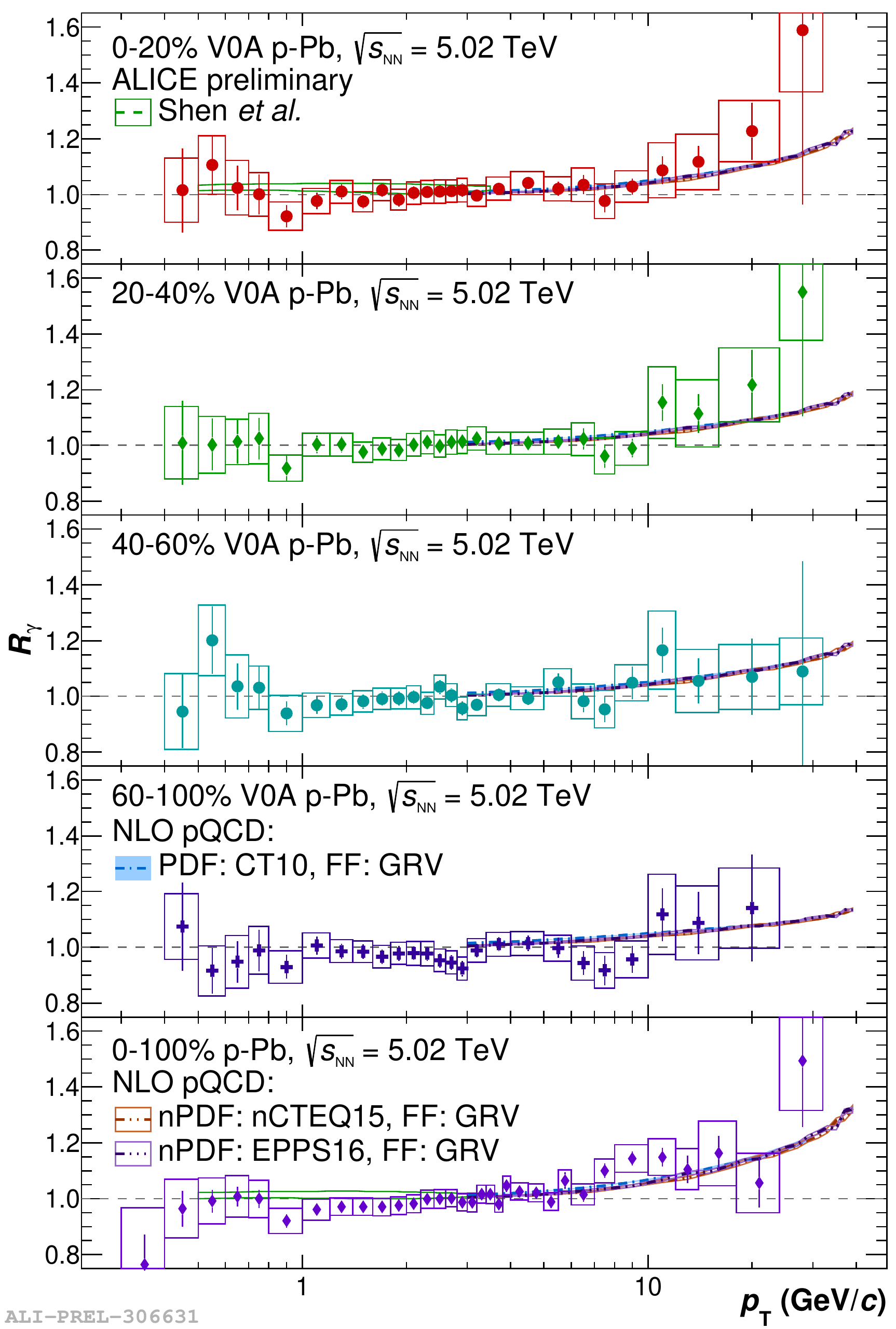}
  \end{center}
  \caption {Direct photon \Rg, labeled by the percentile of the multiplicity distribution.}
\end{wrapfigure}
The Photon Spectrometer PHOS is a homogeneous calorimeter made from $\text{PbWO}_4$ crystals.
The acceptance of $\Delta \varphi = 60^{\circ}$ and $|\eta| < 0.125$ is smaller than the one of EMCal but PHOS has a finer granularity with a cell size of about $2.2 \times 2.2 \,\text{cm}^2$ at $R=4.6$\,m. 
With the photon conversion method (PCM) we exploit the fact that photons can convert to an electron positron pair in the detector 
material with a probability of about $8\%$ (for $R<1.8$\,m).
We search for secondary vertices characterized by two oppositely charged emerging electron tracks which we detect in the Inner Tracking System (ITS, $|\eta| < 1.2$) and Time Projection Chamber (TPC, $|\eta| < 0.9$). 
This methods benefits from the large acceptance and the good momentum resolution especially at low \pT.
In addition to the inclusive photons, we measure \piz and $\eta$ mesons via their decay to two photons, performing an invariant mass analysis of photon pairs. 
For this purpose, the photon samples are taken from the same methods as mentioned above, as well as from one additional hybrid method (PCM-EMC), where one photon is detected with PCM and one with EMCal. The decay photon spectra are then obtained using a Monte Carlo simulation including all relevant hadron decays, based on the measured neutral meson spectra and $m_{\text{\tiny{T}}}$ scaling.
For each of the four methods, the \Rg is calculated. In case of the PCM-EMC method, inclusive photons are taken from PCM.
After checking that the results from all methods are consistent, they were combined. 
The analysis was performed in event multiplicity classes of data from p--Pb collisions recorded in 2013.  
The results are shown in Fig. 1.
The dotted blue, red and purple lines, starting at \pT$ = 3$\,\GeVc  show results of different pQCD calculations.
They are all well compatible with  the measured points at high \pT. 
For 0-20\% and 0-100\% samples a green band is drawn in addition, which shows a prediction from a hydrodynamic model \cite{4} including thermal photon emission at low \pT . 
The current data are not sensitive to the predicted very small thermal photon signal.\\ 


\section{Direct photon \vtwo in  \sNNF Pb--Pb collisions}

The inclusive photon \vtwo is obtained using the scalar product method \cite{5}.
Reference particles are measured in the V0 scintillation detectors placed in a different pseudorapidity region ($2.8 < \eta <5.1$ and $-3.7< \eta <-1.7$) \cite{3}.
Results from the independent methods PCM and PHOS are combined after they were found to be consistent.
In the analysis presented here, Pb--Pb collisions were analyzed in two centrality classes, 0-20\% and 20-40\%.
For PCM, $13.6 \times 10^6$ events were available and $18.8 \times 10^6$ for PHOS.
\begin{figure}
\centering
\includegraphics[width=0.49\textwidth,keepaspectratio=true, trim=0 10 0 20]{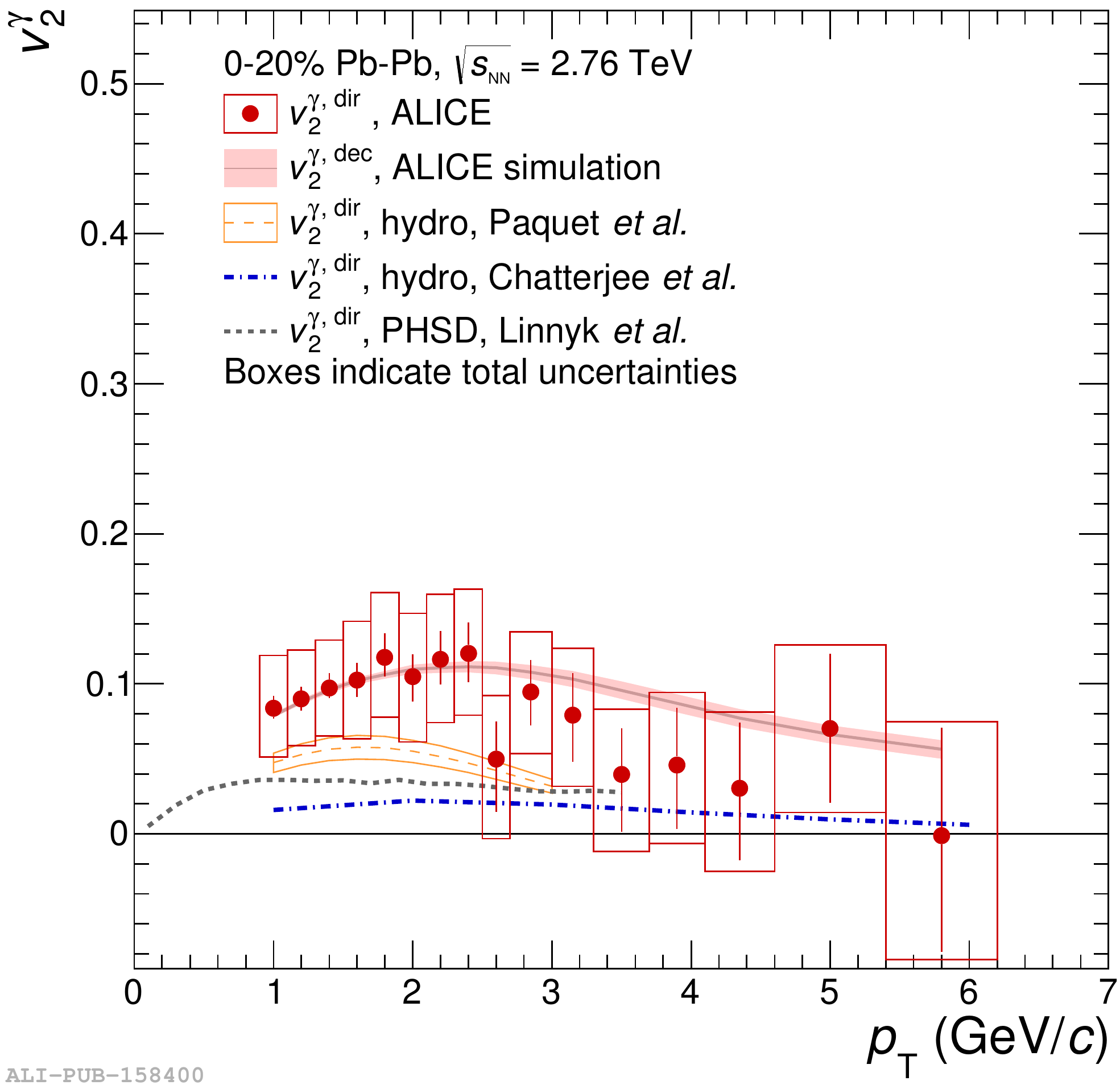}
\includegraphics[width=0.49\textwidth,keepaspectratio=true, trim=0 10 0 20]{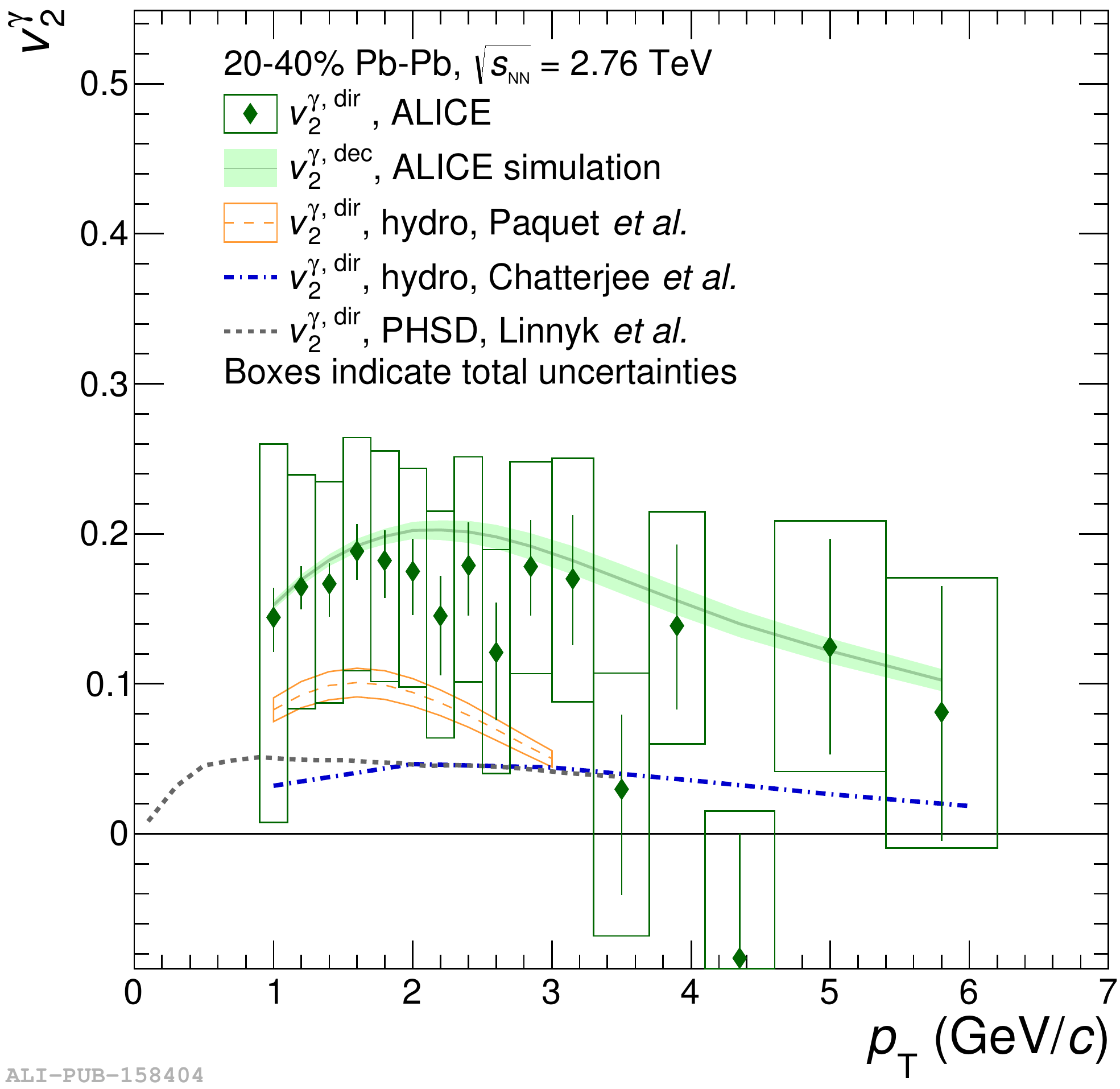}
\caption{Direct and decay photon  \vtwo in central (left) and semicentral (right) collisions compared with hydrodynamic and cascade model predictions \cite{6}.}
\end{figure}
The direct photon $v^{\gamma,\text{dir}}_2$ can be calculated by subtracting the $v^{\gamma,\text{dec}}_2$ of decay photons from the $v^{\gamma,\text{inc}}_2$ of inclusive photons using the following formula:
$v^{\gamma,\text{dir}}_2 = \frac{v^{\gamma,\text{inc}}_2 R_{\gamma} - v^{\gamma,\text{dec}}_2}{R_{\gamma}-1}$,
where the \Rg measured with PCM and PHOS \cite{7} was used.
$v^{\gamma,\text{dec}}_2$ is obtained from a MC simulation including all relevant hadron decays. The simulation is based on hadron spectra and \vtwo measurements and uses $KE_{\text{T}}$ scaling \cite{8} when necessary.
The results for decay photons and direct photons are shown in Fig. 2.
At low \pT, where thermal photons should dominate, we measure a positive $v^{\gamma,\text{dir}}_2$ which is close to $v^{\gamma,\text{dec}}_2$. This indicates an already developed momentum anisotropy of the medium at direct photon production times.
At higher \pT, where the prompt photon contribution increases, the $v^{\gamma,\text{dir}}_2$ decreases. 
In more peripheral events, the thermal photon \vtwo, and therefore also the direct photon \vtwo at low \pT, are expected to be larger than in central events because of the more pronounced initial spatial anisotropy of the medium.
Because the direct photon signal is smaller in more peripheral events \cite{7} the uncertainties are larger in this case and therefore we cannot yet make a conclusive statement on how the direct photon \vtwo changes with centrality.
Calculations from different theoretical models are illustrated by the dashed lines. 
They tend to underestimate the \vtwo with respect to the measured values.


\section{Summary and outlook}

In summary, ALICE has measured the direct photon elliptic flow coefficient \vtwo in Pb--Pb collisions at \sNNF. It was found to be consistent with the current knowledge of the space-time-evolution and photon emission rates but smaller uncertainties will be needed to confirm or exclude given model predictions. 
The direct photon \Rg in high multiplicity p--Pb collisions at \sNNMa is not yet sensitive to the predicted very small thermal photon signal. 
\begin{wrapfigure}[18]{r}{0.45\textwidth}   
  \begin{center} 
  \includegraphics[width=0.45\textwidth,keepaspectratio=true,trim=0 40 0 120]{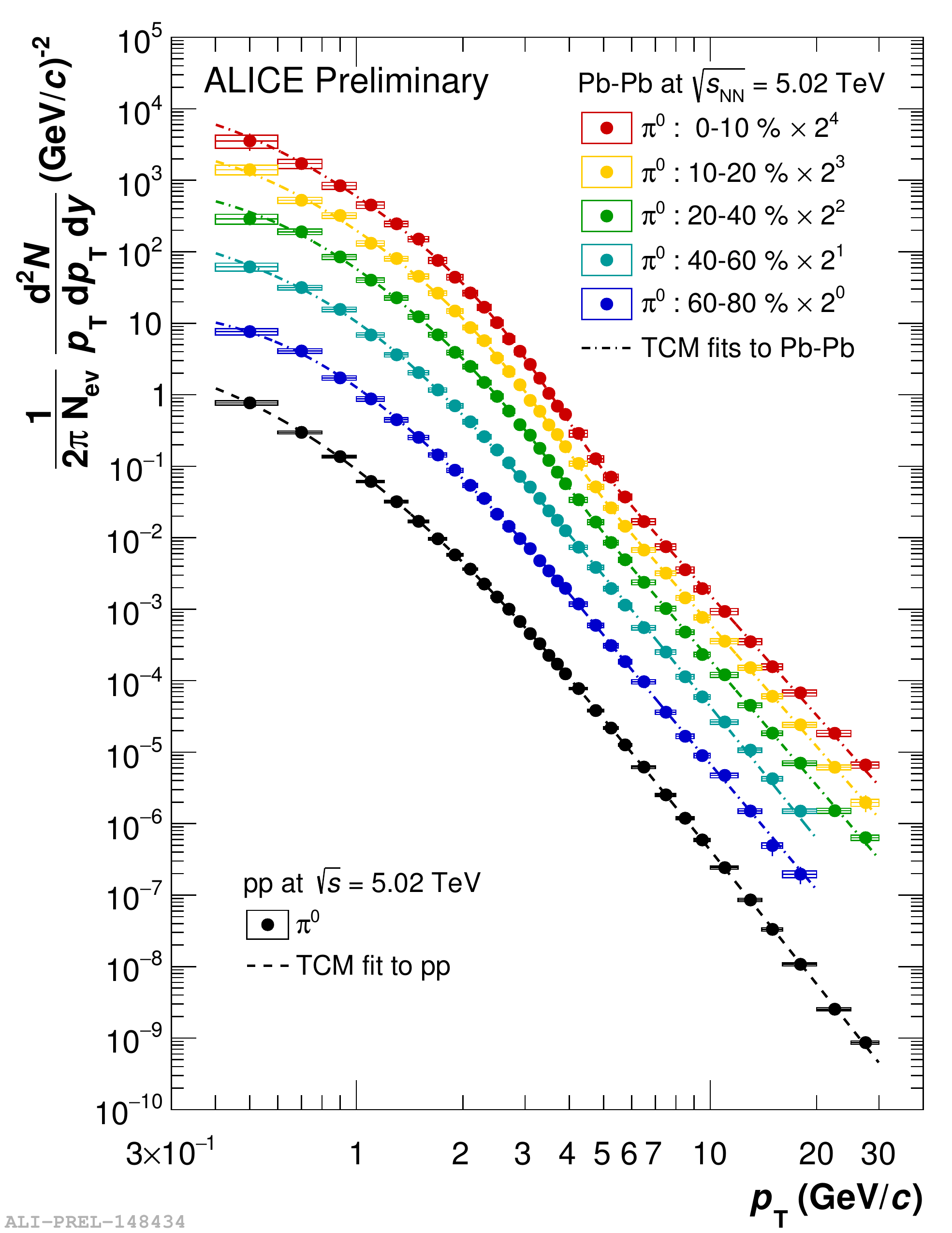}
  \end{center}
  \caption {Neutral pion spectra in pp and Pb--Pb collisions at \sNNMa}
\end{wrapfigure}
Analyses of \sNNMa Pb--Pb collisions are ongoing.
Neutral pion spectra are a crucial input for the decay photon simulation which is required for direct photon measurements.
They are presented in Fig. 3 in different centrality classes. 
The increased statistics of the \sNNMa dataset will enable us to reduce statistical uncertainties of \Rg and \vtwo by one order of magnitude compared to the measurements at  \sNNF. 
It was investigated that also the systematic uncertainties, which dominate over parts of the \pT range, can be reduced by employing new analysis techniques. This allows to exploit the advantages of the ALICE upgrade in future runs.






\end{document}